\documentclass{PoS}

\newcommand{\cV}{\mathcal{V}}
\newcommand{\cH}{\mathcal{H}}

\newcommand{\beqa}{\begin{eqnarray}}
\newcommand{\eeqa}{\end{eqnarray}}
\newcommand{\beq}{\begin{equation}}
\newcommand{\eeq}{\end{equation}}

\title{Infrared behavior of QCD from the Dyson-Schwinger formalism}

\ShortTitle{Infrared behavior of QCD from the Dyson-Schwinger formalism}

\author{{Christian S. Fischer}${}^{ a,b}$
                 \thanks{christian.fischer@physik.tu-darmstadt.de}\\
 \llap{${}^a$} Institut f\"ur Kernphysik, 
  TU Darmstadt, Schlossgartenstra{\ss}e 9, 64289 Darmstadt, Germany\\
  \llap{${}^b$} Gesellschaft f\"ur Schwerionenforschung mbH, 
  Planckstr. 1, 64291 Darmstadt, Germany}

\abstract{We discuss the properties of two different types of 
infrared solutions of Landau gauge Yang-Mills theory and argue 
for one of these (the 'scaling solution'). We furthermore 
clarify the status of previously obtained results from DSEs 
on a four-torus. Including quarks we discuss a relation between 
confinement and dynamical chiral symmetry breaking based on the 
scaling solution of Yang-Mills theory. An infrared singularity 
in the quark-gluon vertex allows for a solution of the $U_A$(1) 
problem along the lines of a mechanism suggested by Kogut and 
Susskind long ago.}

\FullConference{LIGHT CONE 2008 Relativistic Nuclear and Particle Physics\\
                 July 7-11, 2008\\
                 Mulhouse, France}

\begin{document}

\section{Introduction}

There are (at least) two reasons why the Green's functions 
of gauge fixed QCD are interesting objects to study. On
the one hand they are related to fundamental properties of 
the theory like confinement and dynamical chiral symmetry 
breaking. On the other hand they serve as input for 
calculations of observable quantities like dynamical 
properties of bound states, as determined e.g. in the 
framework of Bethe-Salpeter and Faddeev type of equations.

In this talk we are mainly concerned with the first issue.
In the framework of covariantly gauge fixed QCD, Kugo and
Ojima \cite{Kugo:1979gm} have developed a confinement 
scenario that rests on well-defined charges related to 
unbroken global gauge symmetries. In this 
framework BRST-symmetry has been used to identify the positive 
definite space $\cH_{phys}$ of physical states within the total 
state space $\cV$ of QCD. An unbroken global gauge symmetry 
is then crucial to show that the states in $\cH_{phys}$
contributing to the physical S-matrix of QCD are indeed
colorless. They also argued that this setup guarantees the 
disappearance of the 'behind-the-moon' problem, i.e. a 
colorless bound state with colored constituents cannot be 
delocalized into colored lumps \cite{Kugo:1979gm}.

The well-definedness of global gauge symmetry has been 
related to the infrared behavior of the propagators
of Landau gauge QCD in \cite{Kugo:1979gm}: Global gauge 
symmetry is unbroken if in the infrared the ghost propagator 
is more divergent and the gluon propagator less divergent 
than a simple pole. For the gluon propagator this means 
that it is probably at most constant or even vanishing 
in the infrared. In terms of the dressing functions $G(p^2)$ 
and $Z(p^2)$ of the ghost and gluon propagators
\begin{equation}
D_G(p) = -\frac{G(p^2)}{p^2}\,, \hspace*{1cm} 
 D_{\mu \nu}(p) = \left(\delta_{\mu \nu} -
    \frac{p_\mu p_\nu}{p^2}\right) D(p^2) = \left(\delta_{\mu \nu} -
    \frac{p_\mu p_\nu}{p^2}\right)
  \frac{Z(p^2)}{p^2}\,,
 \label{props} 
\end{equation}
and in terms of a power-law expansion this condition reads
\begin{equation}
Z(p^2) \sim (p^2)^{-\kappa_A}; \hspace*{2cm} 
G(p^2) \sim (p^2)^{-\kappa_C}  \label{typeI} 
\end{equation} 
with exponents $\kappa_A \le -1$ and $\kappa_C > 0$.

Nonperturbative information on the ghost and gluon propagators
can be obtained by Dyson-Schwinger equations (DSEs) 
\cite{Alkofer:2000wg} or functional renormalization group 
equations (FRGs) \cite{Pawlowski:2005xe} in the continuum
field theory, or from lattice QCD at finite volume and lattice 
spacing. In the following we first discuss the two possible 
types of numerical solutions (named 'scaling' and 'decoupling')  
in the infinite volume/continuum limit from DSEs. Then we report on
various comparisons between solutions from DSEs on a torus and
results from lattice QCD in section \ref{sec2}. In the last section
we shortly discuss a particular pattern of dynamical chiral symmetry
breaking related to the scaling type of behavior of the Yang-Mills 
sector of QCD.

\section{Infrared Yang-Mills theory from DSEs \label{sec1}}

The infrared behavior of the one-particle irreducible (1PI) Green's 
functions of Yang-Mills theory have been investigated in a number 
of works. The basic relation $\kappa_A=-2\kappa_C$ between the dressing 
functions (\ref{typeI}) of the gluon and ghost propagator 
has been extracted in \cite{von Smekal:1997vx, Lerche:2002ep} 
from DSEs. Corresponding results from FRGs have been obtained in 
\cite{Pawlowski:2003hq}. These findings have been generalized to 
Green's functions with an arbitrary number of legs in 
\cite{Alkofer:2004it}. The analysis rests upon a separation of 
scales, which takes place in the deep infrared momentum region. 
Provided there is only one external momentum 
$p << \Lambda_{\mathrm{QCD}}$ much smaller than $\Lambda_{\mathrm{QCD}}$ 
a self-consistent infrared asymptotic solution of the whole tower of 
Dyson-Schwinger equations for these functions is given by 
\beq
\Gamma^{n,m}(p^2) \sim (p^2)^{(n-m)\kappa}. \label{IRsolution}
\eeq
Here $\Gamma^{n,m}(p^2)$ denotes the dressing function of the infrared 
leading tensor structure of the 1PI-Green's function with $2n$ external 
ghost legs and $m$ external gluon legs. This solution agrees with
the Slavnov-Taylor identities and is the unique scaling solution
in the infrared \cite{Fischer:2006vf}. Here 'scaling' denotes the fact
that {\it all} Green's functions obey nontrivial power laws in the 
infrared with an anomalous dimension $\kappa > 0$ \cite{Watson:2001yv}. 
For the ghost and gluon dressing functions (\ref{typeI}) this scaling 
type of solution yields the abovementioned power law 
$\kappa=\kappa_C=-\kappa_A/2$.

The absence of scaling implies the decoupling of (some) degrees of
freedom. A solution of this type has been discussed e.g. in
\cite{Aguilar:2008xm,Boucaud:2008ji,Dudal:2008sp} and is given by
$\kappa_C = 0$ and $\kappa_A=-1$. We refer to this type of solution 
as the 'decoupling solution'. 
\begin{figure}[b!]
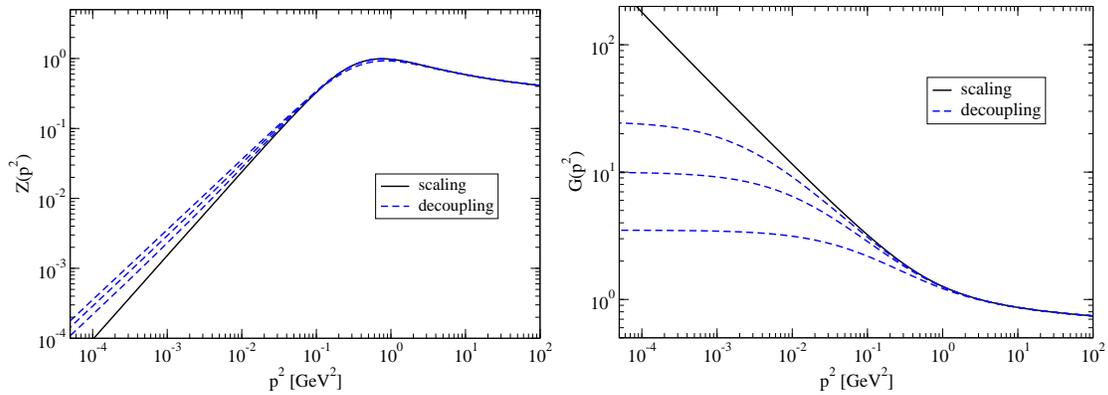

\centerline{\includegraphics[width=0.48\columnwidth]{newglue3.eps}
            \includegraphics[width=0.48\columnwidth]{newghost3.eps}}
  \caption{Numerical solutions for the ghost and gluon
    dressing function with two different boundary conditions $G(0)$.
    The results displayed here are obtained within the truncation
    scheme introduced in \cite{newpaper}. Differences to the scheme defined
    in \cite{Fischer:2002hna} are, however, only very 
    small and would not be visible in the plots.}
\label{resdse}
\end{figure}

Both types of infrared behavior can also be obtained as numerical 
solutions for the coupled systems of ghost and gluon DSEs. In 
\cite{Lerche:2002ep,Boucaud:2008ji} the infrared boundary condition 
$G(0)$, i.e. the value of the ghost dressing function at zero momentum, 
has been identified as a parameter that allows to switch between these 
two types. Clearly, $G(0)^{-1}=0$ 
corresponds to an infrared diverging ghost dressing function implementing
the scaling solution, whereas $G(0)= const.$ produces an infrared finite 
ghost by construction. The gluon propagator is then either massive in the 
sense that $D(0)=\lim_{p^2 \rightarrow 0} Z(p^2)/p^2 = const.$ for 
decoupling, or has the power like behavior (\ref{typeI}) with 
$\kappa=\kappa_C=(93-\sqrt{1201})/98 \approx 0.595353$ \cite{Lerche:2002ep} 
in the case of scaling. The corresponding numerical solutions of the 
coupled ghost and gluon DSEs have been determined in \cite{newpaper} 
and are shown in fig.~\ref{resdse}.

The decoupling type of solution contains an arbitrary and unfixed 
parameter: the value of the ghost at zero momentum and correspondingly 
the finite value $D(0)$ of the gluon. If the gluon were a massive,
physical particle this value could be fixed from experiment. However,
even for decoupling the gluon is {\it not} massive in this sense 
\cite{newpaper} and it is therefore hard to see how $D(0)$ could be 
fixed unambiguously. This problem is absent for the scaling type of 
solutions.  

Although both types of solutions can be obtained from the DSEs, their
status is certainly different. From the discussion in the introduction 
we note that only the scaling type of solutions agrees with the
Kugo-Ojima scenario in the sense that it corresponds to an unbroken 
global gauge symmetry. On the other hand, a broken global gauge symmetry 
is a clear signal for a system in the Higgs phase. We are therefore 
led to the conclusion that the scaling solution represents the confined 
phase of Yang-Mills theory, whereas the decoupling type of solutions 
represents something like a Higgs phase. These arguments and additional
ones related to the breaking of BRST symmetry in the decoupling case 
are discussed in detail in \cite{newpaper}.  

\section{DSEs on a torus: finite volume effects \label{sec2}}

\begin{figure}[b!]
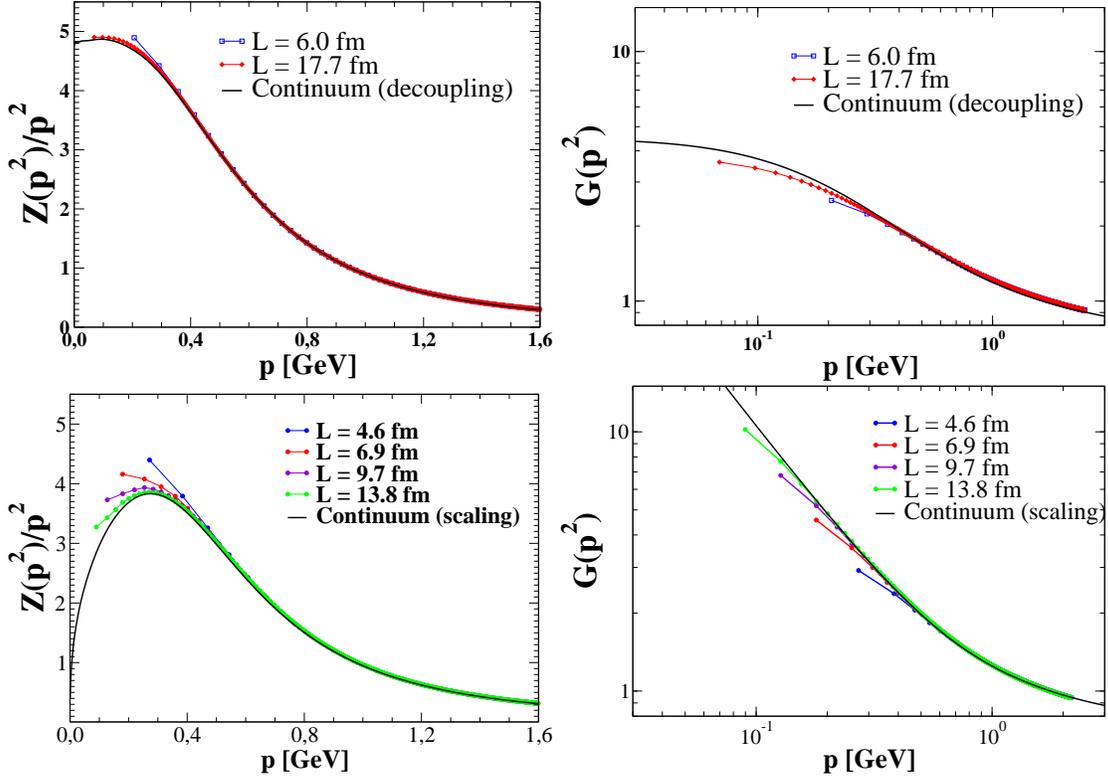

\centerline{\includegraphics[width=0.48\columnwidth]{torus_dec_glue.eps}
            \includegraphics[width=0.48\columnwidth]{torus_dec_ghost.eps}}
\centerline{\includegraphics[width=0.48\columnwidth]{torus_scal_glue.eps}
            \includegraphics[width=0.48\columnwidth]{torus_scal_ghost.eps}}
  \caption{Numerical solutions for the ghost and gluon
    dressing function in the continuum and on tori with different volumes.
    In the upper panel we display solutions of the decoupling type, whereas
    on the lower panel scaling solutions are shown.}
\label{torus}
\end{figure}

In general there are some caveats in comparing results from the 
continuum Dyson-Schwinger approach to those of lattice calculations 
(see \cite{Cucchieri:2007md} and refs. therein). 
The quantitative aspects of the continuum solutions depend on the 
details of the chosen truncation scheme, whereas the lattice 
calculations are {\it ab initio}. Gauge fixing is different in the 
two approaches and the effects of Gribov copies have to be taken 
into account. Furthermore, lattice calculations are carried out on 
a compact manifold, and therefore one has to deal with effects due 
to finite volume and lattice spacing. 

To quantify the 'plain' volume effects (i.e. those not connected to 
the gauge fixing procedure) we formulated the DSEs on a torus without 
changing the truncation scheme. Scaling solutions on a torus have been 
found in \cite{Fischer:2007pf}, whereas solutions of decoupling type 
have been produced in \cite{Fischer:2002eq}. 

In general, one would expect to see differences to the corresponding 
continuum solutions for small volumes, which disappear continuously 
when the volume is chosen larger and larger. This is indeed the case 
as shown in fig.~\ref{torus}. For both types of solutions we obtain 
a smooth infinite volume/continuum limit as the volumes are 
increased\footnote{A corresponding comparison in 
refs.~\cite{Fischer:2002eq,Gruter:2004bb} is misleading since in 
these works decoupling solutions on a torus have been compared 
with the scaling solution in the continuum.}.

It is apparent from the results of fig.~\ref{torus} that volumes 
of $V=(15 \mbox{fm})^4$ and more are necessary to observe signals 
of the infinite volume/continuum behavior of the dressing functions 
also on a torus. As discussed in detail in \cite{Fischer:2007pf} the 
technical reason for this is that one needs a range of momenta $p$ 
with $\frac{\pi}{L} << p << \Lambda_{QCD}$ to observe this behavior; 
these three scales need to be widely separated. In addition it is 
worth noting that the infrared behavior of the Green's functions
does not reflect dynamical properties of the theory. 
These play a role at momenta of the order of or larger than $\Lambda_{QCD}$ 
and are not plagued by volume effects of this magnitude. 
Scaling or decoupling on the other hand are phenomena that occur 
due to the absence of dynamics in the deep infrared momentum region. 
They are characteristic of the global properties of the theory as 
e.g. the conservation or breaking of global gauge symmetries. Scaling 
is also related to the dominance of the Faddeev-Popov determinant 
represented by the ghost degrees of freedom in the infrared. This
dominance allows for the formulation of an infrared effective 
theory where the Yang-Mills part of the Lagrangian can be 
neglected \cite{Zwanziger:2002ia}. 

\section{Dynamical chiral symmetry breaking  \label{sec3}}

In the quark Dyson-Schwinger equation the central object responsible
for dynamical chiral symmetry breaking is the quark-gluon vertex
as the sole carrier of quark-gluon interactions. Based on the scaling 
type of infrared solutions (\ref{IRsolution}), one can derive the 
analytical infrared behavior of this  vertex \cite{Alkofer:2006gz}. 
To this end one has to carefully distinguish the cases of broken or 
unbroken chiral symmetry. Whereas in the broken case the full 
quark-gluon vertex $\Gamma_\mu$ can consist of up to twelve linearly 
independent Dirac tensors, these reduce to a maximum of six when chiral 
symmetry is realized in the Wigner-Weyl mode. Correspondingly, a broken
symmetry induces two tensor structures in the quark propagator, whereas 
only one is left when chiral symmetry is restored. In a similar way, 
chiral symmetry breaking reflects itself in every Green's function 
with quark content.

The presence or absence of the additional tensor structures turns out 
to be crucial for the infrared behavior of the quark-gluon vertex. When 
chiral symmetry is broken (either explicitly or dynamically with a 
valence quark mass $m > \Lambda_{\tt QCD}$) one obtains a self-consistent 
solution of the vertex-DSE which behaves like \cite{Alkofer:2006gz}
\begin{equation}
\lambda^{D\chi SB} \sim (p^2)^{-1/2-\kappa}\,. \label{broken}
\end{equation}
Here $\lambda$ denotes generically any dressing of the twelve tensor 
structures. If, however, chiral symmetry is unbroken one obtains 
the weaker singularity
\begin{equation}
\lambda^{\chi S} \sim (p^2)^{-\kappa}\,. \label{symm}
\end{equation}
As a consequence the running coupling from the quark-gluon vertex either 
is infrared divergent ('infrared slavery') or develops a fixed point:
\begin{equation}
\alpha^{qg}(p^2) = \alpha_\mu \,
{ [\lambda(p^2)]^2} \, { [Z_f(p^2)]^2}\,
{ Z(p^2)} \sim  \left\{ \begin{array}{r@{\quad:\quad}l}
\frac{1}{p^2}\,\,\frac{const_{qg}^{D\chi SB}}{N_c}  & { D\chi SB}\\
\phantom{\frac{1}{p^2}}\,\,\frac{const_{qg}^{\chi S}}{N_c}       &  { \chi S}
\end{array} \right.
\end{equation}
(Here we use that the quark propagator is constant in the infrared, i.e. 
$Z_f(p^2) \sim const$ \cite{Fischer:2003rp}.) Note that in all couplings the
irrational anomalous dimensions ($\sim \kappa$) of the individual dressing 
functions cancel in the RG-invariant products. 

Besides the divergence (\ref{symm}) of the quark-gluon vertex with all momenta
going to zero there also exists a soft collinear-like divergence dependent 
only upon the external gluon momentum $k^2$ \cite{Alkofer:2006gz}:
\begin{equation}
\Gamma \sim \left( k^2 \right)^{-\kappa-1/2}
\label{eqn:softdivergence}
\end{equation}
This additional divergence has two interesting consequences. First, one can 
analyze the behavior of the quark four-point function $H(k^2)$ which includes 
the (static) quark potential. With (\ref{broken}) and (\ref{symm}), one obtains 
$H(k^2) \sim 1/k^4$ in the Nambu-Goldstone and $H(k^2) \sim 1/k^2$ in the 
Wigner-Weyl realization of chiral symmetry. This leads to a quark-antiquark 
potential of
\begin{equation}
V({\bf r}) = \frac{1}{(2\pi)^3} \int d^3k \,\, e^{i {\bf k r}}\, H({\bf k}^2)
\ \ \sim 
\left\{ \begin{array}{r@{\,\,\,:\,\,\,}l}
   { |r|} & { D\chi SB}\\
   { \frac{1}{|r|}}        &  { \chi S}
   \end{array} \right. \,
\end{equation}
which establishes a link between dynamical chiral symmetry breaking
and confinement \cite{Alkofer:2006gz}.

The second consequence concerns the $U_A(1)$-problem.
A confinement driven mechanism for the generation of the topological 
mass of the $\eta'$ in the chiral limit has been suggested by Kogut and Susskind
many years ago \cite{Kogut:1973ab}. It involves the calculation of a certain type
of diagram ('diamond diagram'), which generates such a mass in the presence of 
an infrared divergent gluon propagator $D(k)\sim 1/k^4$ for $k^2\rightarrow0$. 
Today we have excellent evidence that the gluon propagator cannot be that singular.
However, there is the above-mentioned singularity in the quark-gluon vertex. 
Indeed, the combination of a gluon propagator and two dressed vertices appearing 
in the diamond diagram gives precisely a singularity of necessary strength:
\beq
\Gamma(k^2) \frac{Z(k^2)}{k^2} \Gamma(k^2) 
\,\,\,\,\,\,\stackrel{k^2 \rightarrow 0}{\longrightarrow}\,\,\,\,\,\,
(k^2)^{-1/2-\kappa} \,\frac{(k^2)^{2\kappa}}{k^2} \,(k^2)^{-1/2-\kappa} 
\,\,\,\,\,\, \sim \,\,\,\,\,\,1/k^4
\eeq
One then obtains the masses $m_\eta, m_{\eta'}$ and the 
singlet-octet mixing angle $\theta$ of \cite{Alkofer:2008et}
\beq
 \theta = -23.2, \hspace*{1cm} 
 m_\eta = 479 \mbox{MeV}, \hspace*{1cm} 
 m_{\eta'} = 906 \mbox{MeV}
\eeq
in the chiral limit. These values demonstrate that the Kogut-Susskind 
mechanism works in principle. Via the Witten-Veneziano relation one 
obtains the topological susceptibility $\chi^2$ of
\begin{equation}
	\chi^2 = \left( 169\; \textrm{MeV}\right)^4\;,
\end{equation}
in qualitative agreement with lattice results~\cite{DelDebbio:2004ns}.

{\bf Acknowledgments}\\
I am grateful to R.~Alkofer, A.~Maas, F.~Llanes-Estrada, J.~M.~Pawlowski, 
K.~Schwenzer, L.~v.~Smekal and R.~Williams for collaborations on various 
projects related to this talk and to C.~Aguilar, J.~Papavassiliou and O.~Pene 
for useful discussions. This work has been supported by the 
Helmholtz-University Young Investigator Grant number No. VH-NG-332.

\end{document}